\documentclass[twocolumn,prl,showpacs,preprintnumbers,amsmath,amssymb]{revtex4}
\usepackage{CJK}
\usepackage{amsmath}
\usepackage{amssymb}
\usepackage{dcolumn}
\usepackage{graphicx}
\usepackage{bm}

\begin{document}
\begin{CJK}{GB}{gbsn} 
\title{Monte Carlo Study of the Axial Next-Nearest-Neighbor Ising Model}
\author{Kai Zhang}
\affiliation{Department of Chemistry, Duke University, Durham, North
Carolina, 27708, USA}
\author{Patrick Charbonneau}
\affiliation{Department of Chemistry, Duke
University, Durham, North Carolina, 27708, USA}
\date{\today}

\begin{abstract}
The equilibrium phase behavior of microphase-forming systems is notoriously difficult to obtain because of the extended
metastability of the modulated phases. We develop a simulation
method based on thermodynamic integration that surmounts this problem
and with which we describe the modulated regime of the canonical three-dimensional axial
next-nearest-neighbor Ising model. \emph{Equilibrium} order parameters 
are obtained and
the critical behavior 
beyond the Lifshitz point is examined. The absence of widely extended bulging
modulated phases illustrates the limitations of various approximation
schemes used to analyze microphase-forming models.
\end{abstract}
\maketitle
\end{CJK}

Microphases self-assemble in systems with competing short-range attractive and long-range repulsive interactions, irrespective of the physical and chemical nature of these interactions~\cite{seul:1995}. Microphases are the frustrated equivalent of gas-liquid coexistence for purely attracting particles. Periodic lamellae, cylinders, clusters, etc. are thus observed in a variety of systems, such as multiblock copolymers~\cite{hamley:1998}, oil-water surfactant mixtures~\cite{wu:1992}, charged colloidal suspensions~\cite{stradner:2003}, and magnetic materials~\cite{rossat-mignod:1980}. 
Although the modulated organization is spontaneous, obtaining detailed morphological control is notoriously difficult. Annealing~\cite{leung:1986}, external fields~\cite{koppi:1993}, or complex chemical environments~\cite{meli:2009} are usually necessary to order diblock copolymers. Mesoscale periodic textures have found some technological success as thermoplastic elastomers~\cite{hamley:1998} and nanostructure templates~\cite{thurn:2000}, but understanding how to tune and stabilize microphases is essential to broadening their material relevance.

Because experimental systems provide only limited microscopic insight into microphase formation, a number of lattice~\cite{selke:1988,widom:1986,fried:1991,tarjus:2001} and free-space~\cite{sear:1999,cao:2006,archer:2007} models have been put forward. Grasping the equilibrium properties of these models is necessary to resolve the problems surrounding the non-equilibrium assembly of microphases~\cite{cates:1989,charbonneau:2007,toledano:2009}.
Though the modulated regime is a central feature of these
systems, microphases have not been accurately characterized in any
of them. Even for simple models, approximate theoretical frameworks
offer only limited assistance, and treating microphases with
computer simulations is so far an unresolved
problem~\cite{micka:1995,landau:2000}.  In this Letter, we overcome
this last issue by developing a free-energy integration method
for modulated phases. We use this method to determine the phase
diagram of the microphase-forming three-dimensional (3D) axial
next-nearest-neighbor Ising (ANNNI) model, which has reached
textbook status~\cite{chaikin:1995,landau:2000}, but whose characteristic modulated
behavior is still not completely understood. The resulting phase information
allows us to assess the validity of competing approximate treatments
and to better understand the phenomenology of related experimental
systems.

The ANNNI model was introduced nearly half a century ago to explain ``helical'' magnetic order in heavy rare-earth metals~\cite{elliott:1961,selke:1988,yeomans:1988,selke:1992}. Its Hamiltonian on a simple cubic lattice for spin variables $s_i=\pm 1$
\begin{equation}
H_{\mathrm{ANNNI}} =-J\sum_{\langle i,j\rangle}s_{i}s_{j}+\kappa J\sum_{[i,j]}s_{i}s_{j},
\end{equation}
favors alignment of nearest-neighbor pairs $\langle i,j\rangle$, but frustrates long-range order with relative strength $\kappa>0$ for $z$-axial next-nearest-neighbor pairs $[i,j]$. The coupling constant $J$ determines the temperature $T$ scale with Boltzmann's constant $k_B$ set to unity for convenience. The ANNNI model can only be solved exactly in one dimension~\cite{selke:1979}, but some of its higher-dimensional features are nonetheless well understood. In 3D, the topography of the $T$-$\kappa$ phase diagram involves three regions that join together at a multicritical Lifshitz point~\cite{hornreich:1975}: at high $T$ the system is paramagnetic; at low $T$ and $\kappa$ it is ferromagnetic; at low $T$ and for sufficiently high $\kappa$ modulated layered phases form~\cite{selke:1992}.  The ANNNI paramagnetic-modulated (PM) transition beyond the Lifshitz point is thought to be part of the XY universality class~\cite{garel:1976}. For $\kappa<1/2$ the $T=0$ ground state is ferromagnetic, and for $\kappa>1/2$ it is the layered antiphase (``two-up-two-down''). The sequence of commensurate phases springing from the multiphase point at $T=0$ and $\kappa=1/2$, the structure combination branching
processes at low $T$, and the possible occurrence of incommensurate phases are also noteworthy features of the model~\cite{fisher:1980}. 

In order to detail the phase behavior, approximate theoretical treatments, including high- and low-temperature series expansions~\cite{stanley:1977a,fisher:1981}, mean-field~\cite{jensen:1983,selke:1984}, and other theories~\cite{bak:1980,surda:2004,gendiar:2005} 
have been used. Monte Carlo simulations~\cite{selke:1979,selke:1980,rasmussen:1981,kaski:1985} have  reliably determined the paramagnetic-ferromagnetic transition up to the Lifshitz point~\cite{pleimling:2001}, but accurately locating transitions to and within the modulated regime has remained elusive. 
Even within the subset of periodic phases commensurate with the finite lattice, high free-energy barriers need to be crossed on going from one modulated phase to another. Patterns with a metastable nearby periodicity thus persist for very long times~\cite{selke:1979,rasmussen:1981,yeomans:1988}. Traditional simulation methodologies that facilitate ergodic sampling of phase space by passing over such barriers, notably parallel tempering and cluster moves, are of limited help in microphase-forming systems. Because the equilibrium periodicity varies with temperature, sampling higher temperatures leaves the system in a modulated phase with the wrong periodicity; because of the high free-energy barriers between modulated phases and the lack of simple structural rearrangements for sampling different modulations, the efficiency of cluster moves is limited.

\begin{figure}
\includegraphics[width=\columnwidth]{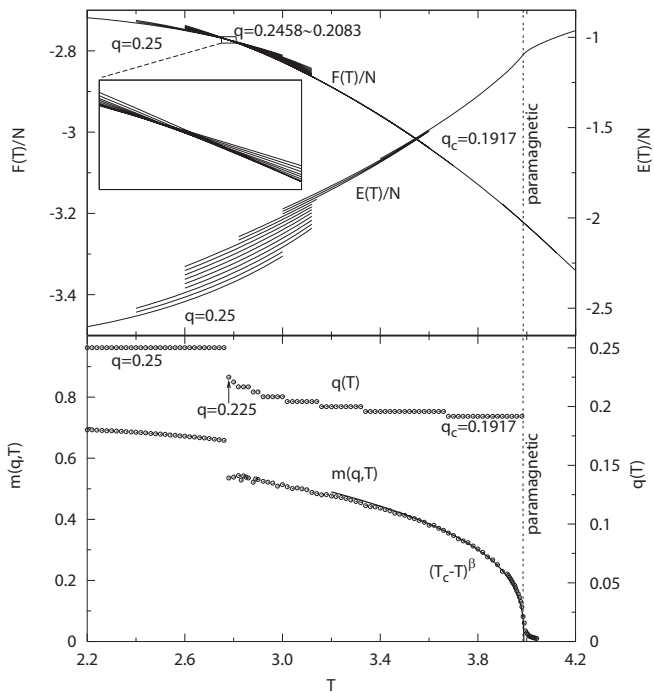}
\caption{(Color online) Simulation results for $\kappa=0.7$. (Top) Energy and free energy per spin for modulations ranging from $q=1/4$ (antiphase)
to $q_c=0.1917$ at melting. The PM transition $T_c=3.988(1)$ (dashed line) is extracted from susceptibility measurements (Fig.~\ref{fig:orderdisorder}). (Bottom) Equilibrium devil's staircase and generalized magnetization $m(q)$. The power-law decay of $m(q)$ with $\beta=0.34(4)$ is superimposed (line). (Inset) Snapshot of the antiphase with differently shaded beads for $s_i=\pm1$.} 
\label{fig:FTANNNI}
\end{figure}

We develop a simulation method based on free-energy integration to treat microphases. The free energy of modulated phases allows us to compare the stability of different periodic patterns and to reliably capture phase transitions. Some aspects of the procedure are part of the standard numerical toolkit~\cite{frenkel:2002}, but additional specifications are in order. For a given $\kappa$, $T$, and wave number modulation $q$, we first calculate the absolute free energy $F$ of $q$-modulated lamellae at a nearby reference temperature $T_0$, and then thermally integrate the energy per spin $E/N$ from $T_0$ to $T$.  
In the spirit of Refs.~\cite{mladek:2007,muller:2008}, the Kirkwood integration begins from decoupled spins under an oscillatory sinusoidal field with Hamiltonian $H_{0}=-B_{0}\sum_{i=1}^Ns_i \sin(2\pi q z_i+\phi_0)$, where a small phase angle $\phi_0$ is added  to prevent the lattice sites from overlapping with the zeros of the field. A scaling field $B_0$ sufficiently strong to avoid melting is necessary for the reversibility of the integration scheme. The high free-energy barriers between the neighboring commensurate periodic patterns would also make phase transitions highly unlikely even if sections of the path are formally metastable~\cite{selke:1979}. Similarly a sinusoidal reference state is valid even if the layer profile squares at low $T$~\cite{selke:1979}, because there is no phase transition along the integration path. We perform constant $T$ Monte Carlo (MC) simulations on a periodic lattice with $N=L_{x} L_{y} L_{z}=40^2\times240$, unless otherwise noted.  Wave numbers $q=n/L_{z}$ for integer values of $n$ keep modulations commensurate with the lattice, which leaves open the problem of incommensurate phases. 
Phase-space sampling gains in efficiency when single-spin flips are complemented with MC moves that take advantage of phase symmetries. In the modulated phases, layer exchanges allow for thickness fluctuations and lattice drifts sample the external field; in the paramagnetic phase, cluster moves accelerate sampling in the critical region~\cite{pleimling:2001}. For $T_0$ reference integrations, up to $10^{5}$ MC moves ($N$ attempted flips) are performed after $5\times10^{4}$ MC moves of preliminary equilibration. 

The smooth and extended energy curves of the different modulations are characteristic of the long-lived metastable nature of the periodic phases (Fig.~\ref{fig:FTANNNI}). Even over relatively long simulation times, metastable systems do not relax to their equilibrium periodicity. Thorough sampling is possible without any modulation change if the $L_x L_y$ cross-section is sufficiently large. The energy gap between neighboring phases for $q$'s commensurate with the simulation box reflects the limited choice of modulations on a finite lattice. In an infinite periodic system, where all rational modulations are valid but irrational $q$'s are excluded, the gap becomes infinitely small because rational numbers are dense on the real axis~\cite{rasmussen:1981}. 
Although they appear to join together smoothly on the scale of Fig.~\ref{fig:FTANNNI}, neighboring free-energy curves intercept. The intercept identifies the transition temperature between two modulated phases with an accuracy that vastly surpasses previous simulation approaches~\cite{selke:1980,rasmussen:1981}. 
Figure~\ref{fig:FTANNNI} illustrates the profile of the
devil's staircase for $\kappa=0.7$~\cite{bak:1980}.
The rate of change of the equilibrium $q$ accelerates upon cooling. The predicted discontinuity of the function before reaching the
antiphase should make the staircase ``harmless'', but
the current numerical accuracy is insufficient to distinguish this
scenario from the ``devil's last step''~\cite{fisher:1987,selke:1988}.

\begin{figure}
\includegraphics[angle=270,width=3.5in]{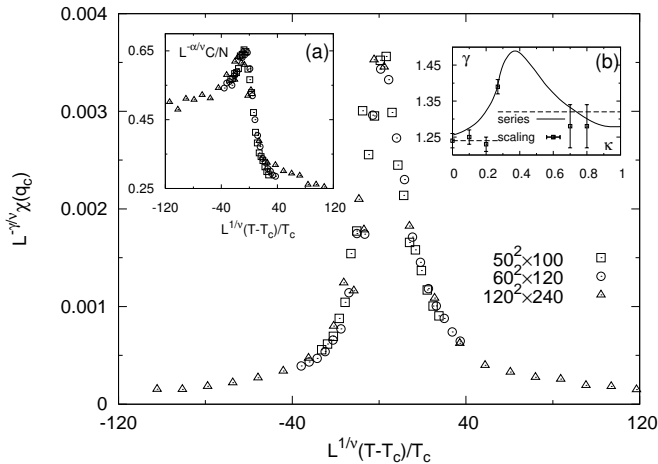}
\caption{Finite-size scaling of $\chi(q_c)$ at $\kappa=0.7$ around the PM transition with $\nu=0.60(3)$ and $\gamma/\nu=2.13(3)$. (a) Same for $C/N$ with $\alpha/\nu=0.18(2)$. (b) Simulation and series expansion $\gamma$ compared with Ising ($\kappa<\kappa_L$) and XY ($\kappa>\kappa_L$) exponents (dashed lines); $\kappa_L=0.270$ result from Ref.~\cite{pleimling:2001}. Rushbrooke and hyperscaling equalities are obeyed within error bars.}
\label{fig:orderdisorder}
\end{figure}
\begin{figure}
\begin{center}
\includegraphics[angle=270,width=3.5in]{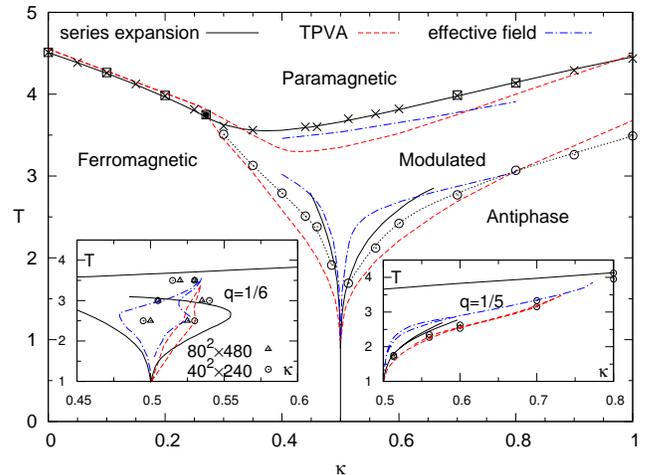}
\caption{(Color online) Lifshitz point ($\bullet$)~\cite{pleimling:2001} and simulation phase boundaries from $\chi(q_{c})$ ($\Box$),
$C$ ($\times$), and $F$ ($\odot$ and dotted line).
High-~\cite{stanley:1977a,oitmaa:1985} and
low-temperature (up to third order around the multiphase point)~\cite{fisher:1981} series expansions as well as
effective-field~\cite{surda:2004} and TPVA~\cite{gendiar:2005}
results are indicated. Stability region of phases
$q=1/6$ (left inset) and $q=1/5$ (right inset) are compared with different theoretical approaches.
} \label{fig:TKfreeenergy}
\end{center}
\end{figure}

The PM critical transition temperature $T_c$, which is analytically well characterized~\cite{stanley:1977a,oitmaa:1985}, is used to validate the simulation results. 
Because the heat capacity per spin $C/N$ is at best only weakly divergent at $T_c$ (Fig.~\ref{fig:orderdisorder}),
we also consider order parameters that are 
functions of the Fourier spin density
$\tilde{s}_q\equiv\sum_{i=1}^{N}s_{i}e^{i 2\pi qz_i}$ and thus naturally capture modulations. In the paramagnetic phase, the
$z$-axis static structure factor
$S(q)\equiv\langle\tilde{s}_q\tilde{s}_{-q}\rangle/N$ grows upon
cooling and diverges at the critical wave number $q_c$ obtained in Fig.~\ref{fig:FTANNNI}~\cite{stanley:1977a,selke:1979}. But the
system-size divergence of $S(q)$ on the modulated side makes it ill-suited for determining $T_c$ in simulations. The generalized magnetization per spin $m(q)\equiv N^{-1}\sqrt{\langle \tilde{s}_{q}\rangle\langle
\tilde{s}_{-q}\rangle}$ also causes problems, because it averages to
zero as the lattice drifts~\cite{tarjus:2001}. To correct for this
problem, we maximize the real component of $\tilde{s}_{q}$ with a
phase shift for each configuration, before taking the thermal average. 
The resulting function shows the characteristic power law
$m(q)\sim|T-T_c|^\beta$ decay (Fig.~\ref{fig:FTANNNI}). The
transition is, however, most clearly identified from the generalized
Binder cumulant~\cite{landau:2000} (not shown) and the generalized
susceptibility
\begin{equation}
N T \chi(q)\equiv\langle \tilde{s}_q\tilde{s}_{-q}\rangle- \langle\tilde{s}_{q}\rangle\langle \tilde{s}_{-q}\rangle=NS(q)-N^2m^2(q),
\label{eq:susceptibility}
\end{equation}
which diverges with system size $\chi(q_c)\sim|T-T_c|^{-\gamma}$
(Fig.~\ref{fig:orderdisorder}) as does $\chi(0)$ at an Ising-like
transition. The $T_c$ results are in very good agreement with the
series expansion~\cite{stanley:1977a,oitmaa:1985}, indicating that
$T_c$ can be identified to within a part in a hundred using $C$ from
the standard lattice size. The resulting determination of the PM
transition (Fig.~\ref{fig:TKfreeenergy}) is also more reliable than
the rare earlier MC results~\cite{selke:1979,rotthaus:1993}, because
of the larger system sizes used.

We also examine the suggested XY character of the PM transition~\cite{garel:1976}. The derivative of $\ln(S(q)/N)$ with $J/T$ gives the correlation length divergence exponent $\nu$, while 
the exponent ratios $\alpha/\nu$ and $\gamma/\nu$ are determined by finite-size scaling of $C/N$ and $\chi(q)$, respectively (Fig.~\ref{fig:orderdisorder})~\cite{landau:2000}.  
The exponent ratios above $\kappa_L$, though consistent with each other, may suggest that the PM transition has a universality that is not of XY type. In particular, $\alpha$ has a positive sign and $\gamma/\nu$ is significantly different from the XY value for the ratio, which may explain the discrepancy of the series expansion $\gamma$ results for large $\kappa$ (see Fig.~\ref{fig:orderdisorder} and caption)~\cite{stanley:1977a,oitmaa:1985}.

More significantly, the approximate treatments, which capture the external boundaries of the modulated regime reasonably well~\cite{surda:2004,gendiar:2005}, 
qualitatively disagree on the internal structure of that regime. On the one hand, as with the mean-field treatment~\cite{jensen:1983,selke:1984} 
and the soliton approximation~\cite{bak:1980}, the effective-field method fills the modulated interior with exceptionally stable ``simple periodic''~\cite{fisher:1987} bulging phases, such as the ``three-up-three-down'' $q=1/6$ phase and the $q=1/5$ phase~\cite{surda:2004}. On the other hand, the tensor product variational approach (TPVA) predicts rather narrow widths for the commensurate phases~\cite{gendiar:2005}. 
The simulation results bulge less than is suggested by the first scenario. The rate of change of $q$ with $\kappa$ and $T$ slows in certain parts of the modulated regime, but all of the phases commensurate with the periodic box are stable in turn. 
The stability range of the different modulated phases is overall fairly small and no exceptional stability is observed for the simple periodic phases $q=1/6$  and $q=1/5$ (Fig.~\ref{fig:TKfreeenergy}), unlike for $q=1/4$. The $q=1/6$ phase does bulge, but increasing the system size, which allows for a more refined $q$ selection, results in a shrinking stability range (Fig.~\ref{fig:TKfreeenergy}), in opposition to the $q=1/4$ phase whose stability range is system-size independent. For the $q=1/5$ phase, the range of stability increases slightly with $\kappa$ in simulation, which is also due to the finiteness of the lattice. 
It is possible that the reduced range of stability of these phases compared to the mean-field predictions be related to the relatively low roughening transition ($T_r= 2.445$~\cite{mon:1990}) in the corresponding Ising model compared to the temperatures studied here. Further study is needed to clarify this point. 
The absence of widely extended bulging phases suggests that the lack of qualitative agreement between observations in magnetic systems, such as
CeSb~\cite{rossat-mignod:1980,muraoka:2002}, and the mean-field stability ranges is to be expected. The commensurate phases observed are those that are kinetically accessible on
experimental time scales~\cite{selke:1979} or whose stability is due to corrections beyond simple spin models. Neither effect suggests a preferable agreement with mean-field predictions.


In this Letter we have presented a methodology for
simulating layered microphases, but modulated assemblies can exhibit
a variety of other symmetries, under the control of an external magnetic field or by tuning the chemical potential in the corresponding lattice gas model. Generalizing the approach to other
order types will greatly benefit the study of more elaborate
microphase-forming systems and pave the way for studies of the non-equilibrium microphase assembly, where most of the materials challenges lie. Generalization to frustrated quantum systems is also conceivable as long as the sign problem can be surmounted~\cite{troyer:2005}.

\begin{acknowledgments}
We thank S. Chandrasekharan, B. Mladek, J. Oitmaa, M. Pleimling, and an anonymous referee. We acknowledge ORAU and Duke startup funding.
\end{acknowledgments}

\end{document}